\documentclass[12pt,a4paper]{article}
\usepackage{graphicx}
\usepackage{soul}
\usepackage{xcolor}
\usepackage[square,numbers,comma]{natbib}
\usepackage{setspace}
\usepackage{fullpage}

\title{ The Hybrid Euclidean-Lorentzian Universe: Stability Considerations and the Point of Transition}

\author{Asher Yahalom\\
Ariel University, Ariel 40700, Israel\\
e-mail: asya@ariel.ac.il
}

\begin{document}
\maketitle
\newcommand{\abs}[1]{ |#1|}
\newcommand{\beq} {\begin{equation}}
\newcommand{\enq} {\end{equation}}
\newcommand{\ber} {\begin {eqnarray}}
\newcommand{\enr} {\end {eqnarray}}
\newcommand{\eq} {equation}
\newcommand{\eqs} {equations }
\newcommand{\mn}  {{\mu \nu}}
\newcommand{\ab}  {{\alpha \beta}}
\newcommand{\abp}  {{\alpha \beta}}
\newcommand{\sn}  {{\sigma \nu}}
\newcommand{\rhm}  {{\rho \mu}}
\newcommand{\sr}  {{\sigma \rho}}
\newcommand{\bh}  {{\bar h}}
\newcommand{\br}  {{\bar r}}
\newcommand {\er}[1] {equation (\ref{#1}) }
\newcommand {\ern}[1] {equation (\ref{#1})}
\newcommand {\Ern}[1] {Equation (\ref{#1})}
\doublespacing
\thispagestyle{empty}

\begin{abstract}
The limited velocity implied by Lorentzian signature, prevents the universe to
reach thermodynamic equilibrium implied by the CMB. Rapid expansion can occur by following Hawking and assuming a primordial Euclidean metric, in which case velocity is not limited and thermalization and rapid expansion are derived without the need to assume an ad-hoc field. Moreover, a mathematical model was given regarding the development of current Lorentzian space-time from the early Euclidean space-time in which problems such as initial singularity and the homogeneity of the CMB are solved. However, stability properties of the hybrid model were not discussed yet, nor was a mathematical criterion for the transition between the Euclidean and Lorentzian era's defined, those will be discussed here to some extent.
\end{abstract}

\vfill \eject
\setcounter{page}{1}

\subsection*{Introduction}

In \cite{PriPar} an extensive introduction to the subject of Euclidean and Lorentzian
 space-times was given, it will not be repeated here, the interested reader is referred to the original paper. We shall only mention that great luminaries of the physical sciences such as Sakharov \cite{Sakharov,Shestakova}, Hawking \cite{Hawking}, Ellis \cite{Ellis} and Davidson \cite{Davidson} have considered the possibility that near its inception the metric of the universe had
 an Euclidean signature. It was shown \cite{Yahaloma} that such a metric cannot be sustained for a long time for an expanding universe since the density of the universe diminishes. The reason is that for a (nearly) empty universes only a Lorentzian metric is stable. So naturally all other metric signature must be  confined to a limited portion of space time. This of course is also true for the Euclidean signature  which is confined as suggested here to a tiny duration after the universe was created. This coupled with the radical dynamics of particles in this primordial Euclidean universe with no upper velocity  limit and bizarre physical statistics that favours high velocities \cite{PriPar}, suggest an alternative for scalar-field driven cosmic inflation \cite{Guth}. A mathematical model for a hybrid Euclidean-Lorentzian universe is described in \cite{Yahalomb}. This is achieved by transforming Einstein equations, taking into account the homogeneity and isotropy of space, into a generalized form of the Friedmann Lema\^{i}tre Robertson Walker (FLRW) equations, which are valid in both the Euclidean and Lorentzian sectors.
 It is shown that within the frame-work of the hybrid Euclidean-Lorentzian universe, problems such as initial singularity and the homogeneity of the CMB are solved. However, stability properties of the hybrid model were not discussed yet, those will be discussed below.

 The structure of the current paper is as follows: we start with a general geometry and apply the standard restrictions that are implied by homogeneity and isotropy, however, we do not restrict our
 metric to be either Lorentzian or Euclidean. This will result in a metric with reduced degrees of freedom. Next we use Einstein equations and obtain the generalized FLRW equations which are valid for both Lorentzian and Euclidean signatures. Finally we discuss the stability of the solutions of those equations, in which we shall attempt to indicate the transition from the Euclidean to the Lorentzian epoch.

\subsection*{The Metric}

The metric $g_\mn$ of a four dimensional space is connected to the infinitesimal square interval $ds^2$ by the well known equation:
\beq
ds^2 = \left| g_\mn dx^\mu dx^\nu \right|, \qquad \mu,\nu \in [0,1,2,3]
\label{dss}
\enq
in which $x^\mu$ are the coordinates describing the location of some point $P$ in this space
and we use the Einstein summation convention.
We shall single out one coordinate $x^0$ and refer to it as "temporal". This can be easily understood in a Lorentzian space-time in which the diagonalize form of $g_\mn$ will have a different sign for this coordinate with respect to the other coordinates. For the Euclidean case the choice seems arbitrary, however, if one bears in mind that any Euclidean portion of space-time will turn into a Lorentzian then the direction of "symmetry breaking" allows us to define a temporal direction. Thus we may write:
 \beq
ds^2 = \left| g_{00} (dx^0)^2 + 2 g_{0k} dx^0 dx^k +
g_{jk} dx^j dx^k \right|. \qquad j,k \in [1,2,3]
\label{dss2}
\enq
Next we invoke isotropy in the usual spatial sense, which is assumed in standard cosmological models to claim that there cannot be a preferred direction in our space at any given time. As the vector
$\vec v$ whose components are $v_k \equiv g_{0k}$ points to such direction, it follows that we must have  $g_{0k} = 0$, and thus:
\beq
ds^2 = \left| g_{00} (dx^0)^2 + g_{jk} dx^j dx^k \right|. \qquad j,k \in [1,2,3]
\label{dss3}
\enq
The next step \cite{Padma} is to look at a "comoving observer", that is an observer that does
not feel himself moving in the coordinate system. Such an observer will report that $dx^k=0$
and thus will be displaced by an interval:
\beq
ds_0^2 = \left| g_{00} (dx^0)^2 \right|.
\label{dss4}
\enq
As the observer is not displaced in space he will interpret the change he is feeling as a
change of time, this is denoted "proper time" $\tau_0$. Till now we have measure all dimensions in the same units (say meters), however, it is customary to measure time using a different set of units
(seconds). To convert between the units we introduce the conversion factor $c \simeq 3~10^8$ m/s (which appears later in the theory as the velocity of gravitational and electromagnetic waves in vacuum). Thus:
\beq
d \tau_0^2 = \frac{ds_0^2}{c^2} = \left| \frac{g_{00}}{c^2} (dx^0)^2 \right|.
\label{dss5}
\enq
There are two alternatives:
\beq
d \tau_0^2 = \left\{
                       \begin{array}{cc}
                         +\frac{g_{00}}{c^2} (dx^0)^2 & g_{00} > 0 \\
                         -\frac{g_{00}}{c^2} (dx^0)^2 & g_{00} < 0 \\
                       \end{array}
                      \right.
\label{dss6}
\enq
So we obtain:
\beq
ds^2 = \left| \pm c^2 d \tau_0^2 + g_{jk} dx^j dx^k \right|.
\label{dss7}
\enq
Now consider an observer who inspects his surroundings in a given instant of time, since space
is assumed to be isotropic he may choose spherical coordinates $r,\theta,\phi$ in which case
\cite{Padma}:
\beq
 g_{jk} dx^j dx^k = A(\tau_0) \left[\lambda^2(r) dr^2 + r^2 (d\theta^2 + \sin^2 \theta d\phi^2 )\right]
\label{dss8}
\enq
As we do not specify in advance the signature of the metric it follows that:
\beq
 A(\tau_0) = \pm a^2 (\tau_0)
\label{dss9}
\enq
Thus we may write the line interval square as:
\beq
ds^2 = \left| \pm c^2 d \tau_0^2 \pm a^2 \left[\lambda^2(r) dr^2 + r^2 (d\theta^2 + \sin^2 \theta d\phi^2 )\right]  \right|.
\label{dss10}
\enq
It follows that there are two equivalent Euclidean choices ($++$ and $--$) and two
equivalent Lorentzian choices ($+-$ and $-+$). Hence without loss of generality we
choose a positive sign for the spatial component, leaving the metric type to be determined by
the temporal part.
\beq
ds^2 = \left| \pm c^2 d \tau_0^2 + a^2 \left[\lambda^2(r) dr^2 + r^2 (d\theta^2 + \sin^2 \theta d\phi^2 )\right]  \right|.
\label{dss11}
\enq
Following \cite{Padma} we shall choose from now on units in which $c=1$ and use the notation:
\beq
dt^2 \equiv \mp   c^2 d \tau_0^2 .
\label{dss12}
\enq
Such that:
\beq
ds^2 = \left| - d t^2 + a^2 \left[\lambda^2(r) dr^2 + r^2 (d\theta^2 + \sin^2 \theta d\phi^2 )\right]  \right|.
\label{dss13}
\enq
This means that for the Euclidean regions of space-time we use an imaginary time coordinate
while for the Lorentzian regions the time coordinate is real:
\beq
t_E = i \tau_0, \quad i = \sqrt{-1}, \qquad t_L = \tau_0.
\label{dss14}
\enq
As the spatial scalar curvature is:
\beq
^3R = \frac{3}{2 a^2 r^3} \frac{d}{d r} \left[r^2 \left(1-\frac{1}{\lambda^2}\right)\right] .
\label{dss15}
\enq
If we assume that space must be homogeneous it follows that the spatial scalar curvature of space
cannot depend on $r$ but of course it can depend on $t$. It follows that:
\beq
r^2 \left(1-\frac{1}{\lambda^2}\right) = c_1  r^4 + c_2
\label{dss16}
\enq
since $\lambda$ is by assumption independent of $t$.
To avoid a singular expression we may choose $c_2 = 0$ and thus obtain:
\beq
\lambda^2  = \frac{1}{1 - c_1 r^2}
\label{dss17}
\enq
It is now customary to redefine $r$ such that $\bar r \equiv r \sqrt{|c_1|}$ for the cases that
$ |c_1| \ne 0$, which will lead to the customary form:
\beq
ds^2 = \left| - d t^2 + \bar a^2 \left[\frac{d \bar r^2}{1-k \bar r^2} + \bar r^2 (d\theta^2 + \sin^2 \theta d\phi^2 )\right]  \right|. ~ \bar a^2  \equiv \frac{a^2}{|c_1|},
 ~ k \in [-1,0,+1]
\label{dss18}
\enq
For the case $ |c_1| = 0$  we take $\bar r \equiv r, \bar a \equiv a$. Finally
we drop the bars for convenience and write:
\beq
ds^2 = \left| - d t^2 + a^2 \left[\frac{d  r^2}{1-k  r^2} +  r^2 (d\theta^2 + \sin^2 \theta d\phi^2 )\right]  \right|.
\label{dss19}
\enq

\subsection*{The FLRW equations}
\label{FLRW}

In the previous section we have went as far as is possible to determine the metric from symmetry
considerations (isotropy \& homogeneity). However, in order to gain more information one must solve Einstein's equations. This will lead to the FLRW model equations:
\beq
 \frac{\dot a^2+k}{a^2} = \frac{8 \pi G}{3} \rho,
\label{Flrweq1}
\enq
\beq
\frac{2 \ddot a }{a} +  \frac{\dot a^2+k}{a^2} = - 8 \pi G p
\label{Flrweq2}
\enq
It is assumed that the pressure $p$ and the energy density $\rho$ are connected through an equation of state and thus given appropriate initial conditions those equations can be integrated. Let us combine \ern{Flrweq1} and \ern{Flrweq2}, this will lead to:
\beq
 \frac{ \ddot a }{a} = - \frac{4 \pi G}{3} (\rho + 3 p),
\label{dda}
\enq
in \cite{Yahalomb} (section 4) we have shown that the above equation implies singularity for a completely Lorentzian universe while it implies a well behaved solution for a hybrid Euclidean-Lorentzian universe.

\subsection*{Stability Analysis}

It is interesting to look at the stability properties of the hybrid Lorentzian-Euclidean universe, in this case one need to look at a new metric:
\beq
g_\mn = g_{(0)~ \mn} + h_\mn
\label{gper}
\enq
in which $g_{(0)~ \mn}$ is FLRW metric defined in \ern{dss19} and $h_\mn$ is a small perturbation.
Inserting this metric into Einstein's equations and keeping only the linear terms  will result in an equation of the form:
\beq
\hat {\cal L}^\ab_\mn (g_{(0)~ \mn}) h_\ab = \delta T_\mn
\label{LinEinseq}
\enq
The operator $\hat {\cal L}$ is a second order differential operator which depends on the FLRW metric, $h_\mn$ has ten degrees of freedom. Following Padmanabhan \cite{Padma} (p. 562) we introduce a conformal time:
\beq
d\eta = \frac{dt}{a}
\label{confortime}
\enq
this time coordinate will be real in the Lorentzian sector but imaginary in the Euclidean sector of the universe. \Ern{dss19} can now be written in the form:
\beq
ds^2 = a^2\left| - d \eta^2 +  \left[\frac{d  r^2}{1-k  r^2} +  r^2 (d\theta^2 + \sin^2 \theta d\phi^2 )\right]  \right|.
\label{dss19b}
\enq
If $k=0$ the above form simplifies to:
\beq
ds^2 = a^2\left| - d \eta^2 +  d \vec x^2  \right|, \qquad \vec x \equiv (x,y,z).
\label{dss19c}
\enq
Following Padmanabhan \cite{Padma} we shall assume for the sake of simplicity that $k=0$, presumably
the value of $k$ will not have significant effect due to the high density of the early universe.
Padmanabhan \cite{Padma} also suggests to partition the perturbation to spatial and temporal
components:
\beq
h_\mn = (h_{00},h_{0j},h_{jk}) = (2 \Phi, w_j, 2 \tilde h_{jk}),
\label{hmnp}
\enq
that is scalar, vector and tensor perturbations. Furthermore, $w_j$ and $\tilde h_{jk}$  are partitioned into curl free and divergence free parts:
\beq
w_j = w_j^\bot + \partial_j \Phi^\|, \qquad \partial^j w_j^\bot = 0.
\label{wj}
\enq
\beq
\tilde h_{jk} =  -\psi \delta_{jk} + \left(\partial_j U_k^\bot + \partial_k U_j^\bot \right)
+\left(\partial_j \partial_k - \frac{1}{3} \delta_{jk} \vec \nabla^2\right) \Phi_1 + h_{jk}^{\bot\bot} .
\label{htjk}
\enq
$\delta_{jk}$ is a Kronecker delta, $U_k^\bot$ is divergence free and thus has two degrees of freedom, $h_{jk}^{\bot\bot}$ is traceless and has zero divergence and thus  has also two degrees of freedom. Padmanabhan \cite{Padma} shows that $h_{jk}^{\bot\bot}$ is also gauge independent. Furthermore, it is shown \cite{Padma} that by a proper choice of gauge one can set $\Phi_1=U_j^\bot =0$. The $h_{jk}^{\bot\bot}$ terms is decoupled from the rest of the perturbation terms.
By using the ansatz:
\beq
h_{jk}^{\bot\bot} = \frac{v(\eta)}{a} e_{jk}^{\bot\bot} e^{i \vec k \cdot \vec x}.
\label{hperjkans}
\enq
in which $e_{jk}^{\bot\bot}$ is constant, and by virtue of \ern{LinEinseq} one obtain the following
equation for $v(\eta)$ (\cite{Padma} p. 566 equation 13.26):
\beq
v'' + \left(|\vec k|^2 - \frac{a''}{a}\right) v = 0, \qquad
v'' \equiv \frac{d^2 v}{d \eta^2}, \quad \quad a'' \equiv \frac{d^2 a}{d \eta^2}.
 \label{veq}
\enq
This implies that there is a critical wave length:
\beq
k_c^2 (\eta)\equiv \frac{a''}{a} =\frac{4 \pi G}{3} a^2 (\rho -3 p)
\label{kc}
\enq
in which we have used \ern{Flrweq1} and \ern{Flrweq2} and assumed $k=0$. It thus follows that
generically speaking we have an equation of the form:
\beq
v'' + \left(|\vec k|^2 - k_c^2 (\eta)\right) v = 0.
 \label{veq2}
\enq
The above equation shows that neither a Euclidean nor a Lorentzian state would be generically stable. In fact for $|\vec k|>k_c$ and thus for (short) wave lengths such that: $\lambda<\lambda_c = \frac{2 \pi}{k_c}$, the Lorentzian metric will prevail and the Euclidean metric will destabilize. While for $\lambda>\lambda_c$ (long) wavelengths the Lorentzian metric will destabilize , and the Euclidean metric will prevail. There are of course special circumstances in which a cosmological stable solution is possible. For example for times for which $k_c^2 (\eta) \leq 0$ the Lorentz solution is the only stable solution. Moreover, even at times in which  $k_c^2 (\eta) > 0$ and the maximal size of the universe is smaller than $\lambda_c$, one will not be able to destabilize the Lorentzian universe with long wave length perturbations. This was also noticed in \cite{PreliminAY} (section 3). On the other hand, if $\lambda_c$ is smaller than the smallest possible scale in which a continuous model is applicable, say at times in which: $l_p > \lambda_c$ in which $l_p$ is the Planck scale one cannot consider in the current context a short wavelength perturbation of the form $\lambda<\lambda_c$ so there is no way to destabilize an Euclidean sector.

We shall adopt the standard density pressure relations ansatz:
\beq
p = w \rho.
\label{eqstat}
\enq
Thus we can write:
\beq
k_c^2 (\eta)=4 \pi G a^2 \rho(\frac{1}{3} -  w)
\label{kc2}
\enq
It follows that in the radiation era, or any other era in which matter is relativistic and thus $w=\frac{1}{3}$ the universe is purely Lorentzian and any Euclidean volume destabilizes. In this situation $k_c=0$ and $\lambda_c = \infty$. In the literature types of matter for which $w>\frac{1}{3}$ also exist, for example
Padmanabhan \cite{Padma} discussed "stiff" matter in which the velocity of sound propagation is equal to the velocity of light in vacuum, in this case $w=1$. Notice, however, that the existence of such types of matter is not verified. For slow moving matter $w=0$, we obtain:
\beq
k_c^2 (\eta) = \frac{4 \pi G}{3} a^2 \rho
\label{kc3}
\enq
Now according to equation (35) of \cite{Yahalomb}:
\beq
\rho  = \rho_c \left(\frac{a}{a_c}\right)^{-3(1+w)} = \rho_c \left(\frac{a}{a_c}\right)^{-3}.
 \label{Flrweq6}
\enq
Hence using equation (57) of \cite{Yahalomb}:
\beq
k_c^2 (\eta) = \frac{4 \pi G}{3 a} a_c^3 \rho_c = \frac{1}{2} \frac{a_c}{a}
\quad \Rightarrow \quad \lambda_c = \frac{2 \pi}{k_c} = 2 \sqrt{2} \pi \sqrt{\frac{a}{a_c}}
\simeq 8.9 \sqrt{\frac{a}{a_c}}.
\label{kc4}
\enq
A typical growing time of such instability is:
\beq
\tau= \frac{a_c \lambda_c}{c}.
\label{kc5}
\enq
If we assume Planck density at $t_c$ (the time of a minimal scale factor, see \cite{Yahalomb} equation (60)), it follows that:
\beq
\tau\approx 1.9 \cdot 10^{-44} ~{\rm seconds}.
\label{kc6}
\enq
but can become larger as the universe grows. Thus the Euclidean universe destabilizes at time scales many magnitudes shorter than suggested by current inflation models. This is presumably followed by a stable Lorentzian solution which requires the thermalization of matter into a relativistic form with
$w=\frac{1}{3}$. The reader is reminded \cite{PriPar} that an Euclidean metric does not restrict matter to move at sub-luminal speeds and thus despite the rapid destabilization of the Euclidean space-time thermalization is indeed achievable.

\subsection*{Conclusion}

We have shown that completely Lorentzian Cosmology described in many textbooks \cite{Padma,Weinberg,MTW,Narlikar} cannot describe the entire history of the universe, as it must breakdown at some time and lead to unphysical divergences. While a Cosmology of a Euclidean and Lorentzian sectors does not contain unphysical singularities and provides sufficient time \cite{PriPar} to the CMB to achieve thermal equilibrium. Indeed black hole singularities can also be avoided in a similar fashion \cite{ArXiv,Geodesic,Particle}. Thus the need to introduce cosmological inflation using ad-hoc (and redundant anywhere else in physics) scalar fields \cite{Guth} is avoided. We mention that the scale factor at the end of the Euclidean epoch serves as an initial
condition for the scale factor in the Lorentzian epoch and this may have some bearing on the apparent scale factor acceleration as evident in the red shift of distant super novae \cite{Yahalomf}. However, some authors suggest to keep the cosmological constant and show how a signature change may affect its value \cite{Bruno}.

The current work of course does not deal with all issues related to the Euclidean Lorentzian transition. Major concerns is related to the neglect of the curvature effect ($k=1$) for the Cosmological model, as well as other types of perturbation. Other concerns are related to the dynamics of particles moving from the Euclidean to the Lorentzian sectors, and the thermodynamics of matter in the Euclidean sector which may be different \cite{PriPar} from the thermodynamics in the Lorentzian sector used in the current paper for both sectors. Those subjects will be dealt with, hopefully, in future works.

\vfill \eject
\newpage

\setcounter{page}{1}

\end{document}